\documentclass[12pt,a4paper]{article}
\usepackage{times}
\usepackage{graphics}
\usepackage{graphicx}
\usepackage{float}
\restylefloat{figure}

\newcommand{\Om}{\mbox{\boldmath $\Omega$}}

\begin{document}
\date{ }

\begin{center}
{\Large  Strongly coupled chameleon fields: possible test with a neutron Lloyd's
mirror interferometer.}


\vskip 0.7cm

Yu. N. Pokotilovski\footnote{e-mail: pokot@nf.jinr.ru}

\vskip 0.7cm
            Joint Institute for Nuclear Research\\
              141980 Dubna, Moscow region, Russia\\
\vskip 0.7cm

{\bf Abstract\\}

\begin{minipage}{130mm}

\vskip 0.7cm
 The consideration is presented of possible neutron Lloyd's mirror
interferometer experiment to search for strongly coupled chameleon fields.
 The chameleon scalar field were proposed to explain the acceleration of
expansion of the Universe.
 The presence of a chameleon field results in a change of a particle's potential
energy in vicinity of a massive body.
 This interaction causes a phase shift of neutron waves in the interferometer.
 The sensitivity of the method is estimated.
\end{minipage}
\end{center}
\vskip 0.3cm


\vskip 0.2cm


\vskip 0.6cm

 There is evidence of the accelerated expansion of the Universe.
 The nature of this effect is one of the most exciting problems in physics and
cosmology.
 Amongst various ideas proposed to explain this astronomical observation
one of popular variants is a new matter component of the Universe -- a
cosmological scalar field of the quintessence type \cite{Ratra} dominating
the present day density of the Universe (for the recent reviews see
\cite{Pee,Cop}).

 Acting on cosmological distances the mass of this field should be of order
of the Hubble constant: $\hbar H_{0}/c^{2}=10^{-33}\,eV/c^{2}$.

 The massless scalar fields appearing in string and supergravity theories
couple to matter with gravitational strength.
 Because of direct coupling to matter with a strength of gravitation existence
of light scalar fields leads to large violation of the equivalence principle.
 In the absence of self-interaction of the scalar field the experimental
constraints on such a field are very strict demanding their coupling to matter
to be unnaturally small.

 The solution proposed in \cite{Kho,Bra04,Gub,Upa06,Mota06,Mota07} consists in
the introduction of the coupling of the scalar field with matter of such a form
that in result of self-interaction and interaction of the scalar field with
matter the mass of the scalar field depends on the local matter environment.

 In the proposed variant coupling to matter is of order as expected
from string theory, but is very small on cosmological scales.
 In surrounding with high matter density the mass of the field is increased,
the interaction range is strongly decreased, and the equivalence principle is
not violated in laboratory experiments for the search for the long range fifth
force.
 The field is confined inside the matter screening its existence to the
external world.
 In this way the chameleon fields evade tests of the equivalence principle and
the fifth force experiments even if these fields are strongly coupled to matter.
 In result of the screening effect the laboratory gravitational experiments are
unable to set an upper limit on the strength of the chameleon-matter coupling.

 The deviations of results of measurements of gravity forces at macroscopic
distances from calculations based on Newtonian physics can be seen in the
experiments of Galileo-, E\"otv\"os- or Cavendish-type \cite{Fisch} performed
with macro-bodies.
 At smaller distances  ($10^{-7}-10^{-2}$) cm the effect of these forces can be
observed in measurements of the Casimir force between closely placed
macro-bodies (for review see \cite{Rep}) or in the atomic force
microscopy experiments.
 Casimir force measurements may evade to some degree the screening and probe
the interactions of the chameleon field at the micrometer range despite the
presence of the screening effect \cite{Mota07,Cas,Cas1}.

 At even smaller distances such experiments are not sensitive enough, and high
precision particle scattering experiments may play their role.
 In view of absence of electric charge the experiments with neutrons are more
sensitive than with charged particles, electromagnetic effects in scattering of
neutrons by nuclei are generally known and can be accounted for with high
precision \cite{myN,NesvN}.

 As for the chameleon interaction of elementary particles with bulk matter, it
was mentioned in \cite{BraPi} that neutron should not show a screening effect -
the chameleon induced interaction potential of bulk matter with neutron can be
observed.
 It was proposed also in \cite{BraPi} to search for chameleon field through
energy shift of ultracold neutrons in vicinity of reflecting horizontal mirror.
 From already performed experiments on observation of gravitational levels of
neutrons the constraints were obtained in \cite{BraPi} on parameters,
characterizing the force of chameleon-matter interaction.

 Chameleons can also couple to photons.
 In \cite{Ahl,Gies} it was proposed to search for the afterglow effect in a
closed vacuum cavity resulting from chameleon-photon interaction in magnetic
field.
 The GammeV-CHASE \cite{CHASE,CHASE1} and ADMX \cite{ADMX} experiments based on
this approach are intended to measure (constrain) the coupling of chameleon scalar field to matter and photons.

 In the approach proposed here only chameleon-matter interaction is measured
not relying on existence of the chameleon-photon interaction.
 It is based on the standard method of measurement phase shift of a neutron
wave in the chameleon-neutron interaction potential.

 Irrespective of any particular variant of the theory, test of the interaction
of particles with matter at small distances may be interesting.

 In one of popular variants of the chameleon scalar field theory
\cite{Kho,Bra04,Gub,Upa06,Mota06,Mota07} the chameleon effective potential is
\begin{equation}
V_{eff}(\phi)=V(\phi)+e^{\beta\phi/M_{Pl}}\rho,
\end{equation}
where $V(\phi)$ is the scalar field potential:
\begin{equation}
V(\phi)=\Lambda^{4}+\frac{\Lambda^{4+n}}{\phi^{n}},
\end{equation}
and $\rho$ is the local energy density of the environment.
 In these expressions $\Lambda=(\hbar^{3}c^{3}\rho_{d.e.})^{1/4}$=2.4 meV is
the dark energy scale, $\rho_{d.e.}\approx 0.7\times 10^{-8}$ erg/cm$^{3}$ is
the dark energy density.

 The chameleon interaction potential of a neutron with bulk matter (mirror)
was calculated in \cite{BraPi}:
\begin{equation}
V(z)=\beta\frac{m}{M_{Pl}\lambda}\Bigl(\frac{2+n}{\sqrt{2}}\Bigr)^{2/(2+n)}
\Bigl(\frac{z}{\lambda}\Bigr)^{2/(2+n)}= \beta\cdot 0.9\cdot 10^{-21}\, eV
\Bigl(\frac{2+n}{\sqrt{2}}\Bigr)^{2/(2+n)}
\Bigl(\frac{z}{\lambda}\Bigr)^{2/(2+n)}=
{V_{0}}\Bigl(\frac{z}{\lambda}\Bigr)^{2/(2+n)},
\end{equation}
\begin{equation}
V_{0}=\beta\cdot 0.9\cdot 10^{-21}\,eV
\Bigl(\frac{2+n}{\sqrt{2}}\Bigr)^{2/(2+n)},
\end{equation}
here $\lambda=\hbar c/\Lambda=82\,\mu m$.


 The Lloyd's mirror neutron interferometer can provide high sensitivity to
the chameleon forces in the large $\beta$ range.

 Fig. 1 shows the geometry of the Lloyd's mirror interferometer.
 To decrease strong effect of the Earth gravity the mirror of the
interferometer is vertical.

 The neutron wave vector $k^{'}$ in the potential $V$ is
\begin{equation}
k^{'2}=k^{2}-\frac{2mV}{\hbar^{2}}, \qquad k^{'}=k-\frac{mV}{k\hbar^{2}},
\end{equation}
where $m$ is the neutron mass and $k$ is the neutron wave vector in the absence
of any potential.

 The phase shift due to the chameleon mediated interaction potential of a
neutron with the mirror,
depending on distance from the mirror, is obtained by integration along
trajectories $\varphi=\oint k^{'}ds=\varphi_{II}-\varphi_{I}$, where
$\varphi_{I}$ and $\varphi_{II}$ are the phases obtained along trajectories I
and II respectively:
\begin{eqnarray}
\varphi_{I} & = & k\sqrt{L^{2}+(b-a)^{2}}-
\frac{mV_{0}\sqrt{1+((b-a)/L)^{2}}}{k\hbar^{2}\lambda^{\alpha_{n}-1}}
\int_0^L(a+\frac{b-a}{L}x)^{\alpha_{n}-1}dx= \nonumber\\
& & = \varphi_{I,geom}-\frac{\gamma\sqrt{1+((b-a)/L)^{2}}}
{\lambda^{\alpha_{n}-1}\alpha_{n}(b-a)}
\Bigl(b^{\alpha_{n}}-a^{\alpha_{n}}\Bigr)
\end{eqnarray}
and
\begin{eqnarray}
\varphi_{II} & = & k\sqrt{L^{2}+(b+a)^{2}}-
\frac{mV_{0}\sqrt{1+((b+a)/L)^{2}}}{k\hbar^{2}\lambda^{\alpha_{n}-1}}
\Biggl[\int_0^l(a-\frac{b+a}{L}x)^{\alpha_{n}-1}dx
+\int_l^L(\frac{b+a}{L}x-a)^{\alpha_{n}-1}dx\Biggr]= \nonumber\\
& & = \varphi_{II,geom}+\frac{\gamma\sqrt{1+((b+a)/L)^{2}}}
{\lambda^{\alpha_{n}-1}\alpha_{n}(b+a)}
\Bigl(b^{\alpha_{n}}+a^{\alpha_{n}}\Bigr).
\end{eqnarray}

 Here $l=(aL)/(a+b)$ is the x-coordinate of the beam II reflection point from
the mirror, $\gamma=(mV_{0}L)/(k\hbar^{2})$, and $\alpha_{n}=(4+n)/(2+n)$.

 The geometric phase shift
\begin{equation}
\varphi_{geom}=\varphi_{II,geom}-\varphi_{I,geom}=
k\Bigl(\sqrt{L^{2}+(b+a)^{2}}-\sqrt{L^{2}+(b-a)^{2}}\Bigr)\approx 2kab/L
\end{equation}
with relative precision better than $ab/L^{2}$.
 The geometric phase shift linearly depends on position of the interference
coordinate $b$.
 It means that the interference pattern in absence of any potentials is
sinusoidal with high precision:  $ab/L^{2}\sim 10^{-8}$ at
$a\sim b\sim 10^{-2}$\,cm and $L=100$\,cm.

 The phase shift from the chameleon neutron-mirror potential
\begin{eqnarray}
\varphi_{cham} & = & \varphi_{II,cham}-\varphi_{I,cham}= \nonumber\\
& & = \frac{\gamma}{\lambda^{\alpha_{n}-1}\alpha_{n}}
\Bigl[\frac{b^{\alpha_{n}}-a^{\alpha_{n}}}{b-a}\sqrt{1+((b-a)/L)^{2}}-
\frac{b^{\alpha_{n}}+a^{\alpha_{n}}}{b+a}\sqrt{1+((b+a)/L)^{2}}\Bigr]
\approx \nonumber\\
& & \approx \frac{\gamma}{\lambda^{\alpha_{n}-1}
\alpha_{n}}2ab \frac{b^{\alpha_{n}-1}-a^{\alpha_{n}-1}}{b^{2}-a^{2}}.
\end{eqnarray}

 For non-strictly vertical mirror the component of the Earth gravity normal to the
surface of the mirror produces the potential $V_{gr}=cmgz$, where $g$ is
the gravitational acceleration, and the coefficient $c$ depends on the angle
$\theta$ between gravity vector and the mirror plane.
 At $\theta=10^{"}$, $c\approx 5\times 10^{-5}$.
 This linear potential leads to additional phase shift
\begin{equation}
\varphi_{gr}=\varphi_{II,gr}-\varphi_{I,gr}=\frac{cgm^{2}}{2k\hbar^{2}(a+b)}
\Bigl[(b+a)^{2}\sqrt{L^{2}+(b-a)^{2}}-(b^{2}+a^{2})\sqrt{L^{2}+(b+a)^{2}}\Bigr]
\approx \frac{cgm^{2}}{k\hbar^{2}}\frac{abL}{a+b},
\end{equation}
calculated in analogy with Eqs.(6-9).

 The Coriolis phase shift due to the Earth rotation \cite{Cori} is
\begin{equation}
\varphi_{Cor}=\frac{2m}{\hbar}(\Om\bf A),
\end{equation}
where $\Om$ is the vector of angular rotation of the Earth and $\bf A$
is the vector of the area enclosed by the interferometer's beams.

 As $A=(abL)/(2(a+b))$, for the Institute Laue-Langevin geographic position
(where good very cold neutron source has been constructed \cite{Ste}) we have
$\varphi_{Cor}=0.16 (abL)/(2(a+b))$\,rad (a,b,L in cm).
 As expected it is similar to the gravitational phase shift in its
dependence of the slit and the interference coordinates.

 We should calculate also the phase shift of the neutron wave along the
beam II at the point of reflection.
 Neglecting imaginary part of the potential of the mirror, the amplitude of
the reflected wave is $r=e^{-i\varphi_{refl}}$, with the phase
\begin{equation}
\varphi_{refl}=2\arccos(k_{norm}/k_{b})\approx\pi-\delta\varphi_{refl},
\end{equation}
where
\begin{equation}
\delta\varphi_{refl}=2k_{norm}/k_{b}\approx\pi-2\frac{k}{k_{b}}\frac{a+b}{L}.
\end{equation}
 Here $k_{norm}$ is normal to the mirror's surface component of the neutron
wave vector, and $k_{b}$ is the boundary wave vector of the mirror.
 This phase shift linearly depends on $b$ similarly to the
geometric phase shift $\varphi_{geom}$.

 The reflected and non-reflected beams follow slightly different paths in the
interferometer.
 Therefore in the vertical arrangement of the reflecting mirror they spent
different times in the Earth's gravitational field $\Delta t=2ab/(Lv)$.
 The difference in vertical shifts of the reflected and non-reflected beams is
$\Delta h=2gab/v^{2}$, and the phase shift due to this difference
\begin{equation}
\Delta\varphi_{vert}=kg^{2}abL/v^{4}.
\end{equation}
 At our parameters of the interferometer this value is of order $\sim 10^{-4}$.

 The total measured phase shift is
\begin{equation}
\varphi=\varphi_{geom}+\varphi_{cham}+\varphi_{gr}+\varphi_{Cor}+\varphi_{refl}.
\end{equation}

 The gravitational phase shift can be suppressed by installing the mirror
vertically with highest possible precision.
 On the other hand the gravitational phase shift may be used for calibration of the
the interferometer by rotation around horizontal axis.

 The phase shifts due to the Earth rotation $\varphi_{Cor}$ and reflection
$\varphi_{refl}$ may be calculated and taken into account in analysis of the interference curve.

 Figs. 2 shows the calculated phase shift $\varphi_{cham}$ for an idealized Lloyd's
mirror interferometer (strictly monochromatic neutrons, width of the slit
is zero, detector resolution is perfect) with parameters: $L$=1 m,
$a=100\,\mu$m, the neutron wave length $\lambda_{n}$=100 \AA,
(the neutron velocity 40 m/s), $\beta=10^{7}$, at $n$=1 and $n$=6.
 Shown also are the gravitational phase shift $\varphi_{gr}$ at
$c=5\times 10^{-5}$, and $\delta\varphi_{refl}$ = $\pi-$phase shift of the
ray II at reflection ($k_{b}=10^{6}$\,cm$^{-1}$).

 It is essential that searched for the phase shift due to hypothetical
chameleon potential depends on the interference coordinate nonlinearly and has
to be inferred from analysis of the interference pattern.

 Fig. 3 demonstrates the calculated interference pattern for the same parameters
of the interferometer: 1 - all the phase shifts are taken into account: of the chameleon field with matter
interaction parameters $\beta=10^{7}$ and $n=1$, the gravitational phase shift $\varphi_{gr}$ at $c=10^{-4}$,
the Coriolis phase shift, and the phase shift of the ray II at reflection
$\delta\varphi_{refl}=\pi-\varphi_{refl}$ ($k_{b}=10^{6}$\,cm$^{-1}$); 2 - the same, but excluding all the phase
shifts except the geometrical and the chameleon induced phase shifts; 3 - all phase shifts excluded except
purely geometrical. In the latter case the interference pattern should be strongly sinusoidal with period of
oscillations determined by the geometric phase shift: $\Lambda=\lambda_{n}L/(2a)$.
 The number of oscillations in an interference pattern with the coordinate less
than $b$ is $n_{osc}=2ab/(\lambda_{n}L)$.

 The spatial resolution of the position sensitive slow neutron detectors is at
the level 5 $\mu$m with electronic registration \cite{eldet} and about 1 $\mu$m
with the plastic nuclear track detection technique \cite{emdet}.

 It follows from these calculations that the effect of the chameleon
interaction of a neutron with matter may be tested in the range of strong
coupling at the parameter of interaction down to $\beta\sim 10^{7}$ or lower.

 Existing constraints on the parameters $\beta$ and $n$ may be found in Fig. 1
of Ref. \cite{BraPi}.
 For example the allowed range of parameters for the strong coupling regime
$\beta\gg 1$ are: $50<\beta<5\times 10^{10}$ for n=1,
$10<\beta<2\times 10^{10}$ for n=2, and $\beta<10^{10}$ for $n>2$.
 It is seen that the Lloyd's mirror interferometer may be able to constrain
the chameleon field in the large coupling area of the theory parameters.

 From \cite{Ste} where the phase density at the PF-2 VCN channel in ILL was
measured to be 0.25 cm$^{-3}$ (m/s)$^{-3}$ at v=50 m/s it is possible to
estimate the VCN flux density:
 $\phi_{VCN}=9.3\times 10^{4}$ cm$^{-2}$s$^{-1}$\AA$^{-1}$.
 On the other hand \cite{Dre} gives for the same channel larger value:
$\phi_{VCN}=4\times 10^{5}$ cm$^{-2}$s$^{-1}$(m/s)$^{-1}$.

 Using the Zernike theorem it is possible to calculate the width $d$ of the
slit necessary to satisfy good coherence within the coherence aperture
$\omega$, i.e. the maximum angle between diverging interfering beams:
$x=\pi \omega d/\lambda_{n}\leq 1$, if to irradiate the slit with an incoherent
flux.
 As $\omega=2b/L$, $d\leq L\lambda_{n}/(2\pi b)\sim 2 \mu$m at $b=1$ mm
(20 orders of interference at the period of interference
$\Lambda=\lambda_{n}L/(2a)=50 \mu$m at $a=100 \mu$m).

 At the monochromaticity 5 \AA (the coherence length $\l_{coh}=20\lambda_{n}$),
the slit width 2 $\mu$m, the length of the slit 3 cm, the divergence of the
incident beam determined by the boundary velocity of the VCN guide 7 m/s:
$\Omega=7/100=0.07$, the interference aperture
$\omega=2b/L=0.2/100=2\times 10^{-3}$, so that $\omega/\Omega=0.03$,
we have the count rate to all interference curve with a width of 1 mm
(20 orders of interference):
$10^{5}\times 5\times 2\cdot 10^{-4}\times 3\times 0.03\sim 10$ s$^{-1}$.
 In a one day measurement the number of events in one period of interference
is $\sim 10^{5}$.
 It is more than enough to measure the phase shift of $\sim 0.1$ corresponding
to the effect at $\beta=10^{7}$.

 In the interferometers of the LLL-type \cite{BH,RTB} an interference pattern
is obtained point by point by rotation of a phase flag introduced into the
beams.
 In case of the VCN three grating interferometers (for example
\cite{Io,Eder,Zo}) the phase shift between the beams is realized by the same
method or by shifting position of the grating.
 Distinctive feature of the Lloyd's mirror interferometer is the possibility to
register all the interference pattern simultaneously along z-coordinate
starting from z=0.
 The measured  interference pattern is then analyzed on the subject of
presence of the searched for effects, after corrections on known gravity,
Coriolis, and reflection phase shifts.

 The experiment may be performed with monochromatic very cold neutrons, or in
the time-of-flight mode using large wave length range, for example 80-120 \AA.
 The pseudo-random modulation \cite{corr1,corr2} is used in the correlation
time-of-flight spectrometry.
 It was realized in the very low neutron energy range \cite{corr3}.
 In this case a two-dimensional: interference coordinate -- time-of-flight
registration gives significant statistical gain.

  As in the VCN interferometers based on three gratings in the Lloyd's mirror
neutron interferometer the space between the beams is small (parts of mm).
 Therefore it hardly can be used in experiments where some devices are
introduced in the beams, or between the beams (for example to investigate
non-local quantum-mechanical effects).
 But it may be applicable to search for short-range interactions when they are
produced by reflecting mirror.

 We may estimate sensitivity of the LLL-type interferometer to the chameleon
potential.
 The phase shift in this case is
\begin{equation}
\varphi_{LLL}=\frac{2\gamma\sqrt{1+(2a/L)^{2}}}{\lambda^{\alpha_{n}-1}}
(2a)^{\alpha_{n}-1},
\end{equation}
 where $a$ is the half distance between the beams of the LLL-interferometer.
 The sensitivity of the Lloyd's mirror and the LLL-interferometers is
determined by the factor $La^{\alpha_{n}-1}/k$, which is much in favor of the
Lloyd's mirror interferometer.

 There are strong demands to flatness and waviness of the reflecting mirror:
waviness of $10^{-6}$ rad causes spreading of the interference
picture of $5\times 10^{3}$ rad at the parameters of interferometer
of Fig. 2.

 Significant potential to measure possible new neutron-matter interactions has the recently tested
resonance method of measuring the difference between the neutron gravitational levels \cite{res}.
 Sensitivity of this method depends on the resonance width determined by the neutron
flight time above the mirror.
 In \cite{res} this time was about 15 -- 25 ms.
 To increase this time neutrons should be stored above the mirror in closed cavity, which will
strongly decrease the measured intensity.

 Previously the capability of the Lloyd's mirror neutron interferometer for
the search for hypothetical axion-like spin-dependent forces was considered in
\cite{ax-Lloyd} and for the search for short-range non-Newtonian gravity in
\cite{Newt}.


\newpage
\begin{figure}
\begin{center}
\resizebox{16cm}{9cm}{\includegraphics[width=\columnwidth]{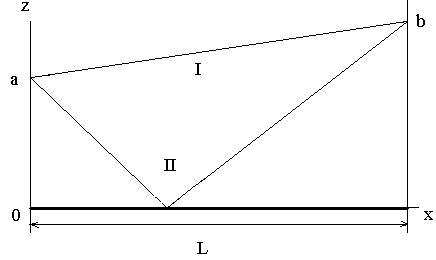}}
\end{center}
\caption{Lloyd's mirror neutron interferometer.
 Reflecting plane is vertical so that the gravity effect on the interference is
strongly reduced.
 The position of the slit in respect to the reflecting plane is $a$, $L$ is the
distance from the slit to the detector plane, $b$ is the distance of the
detector coordinate from the reflecting mirror.}
\end{figure}

\newpage

\begin{figure}
\begin{center}
\resizebox{16cm}{13cm}{\includegraphics[width=\columnwidth]{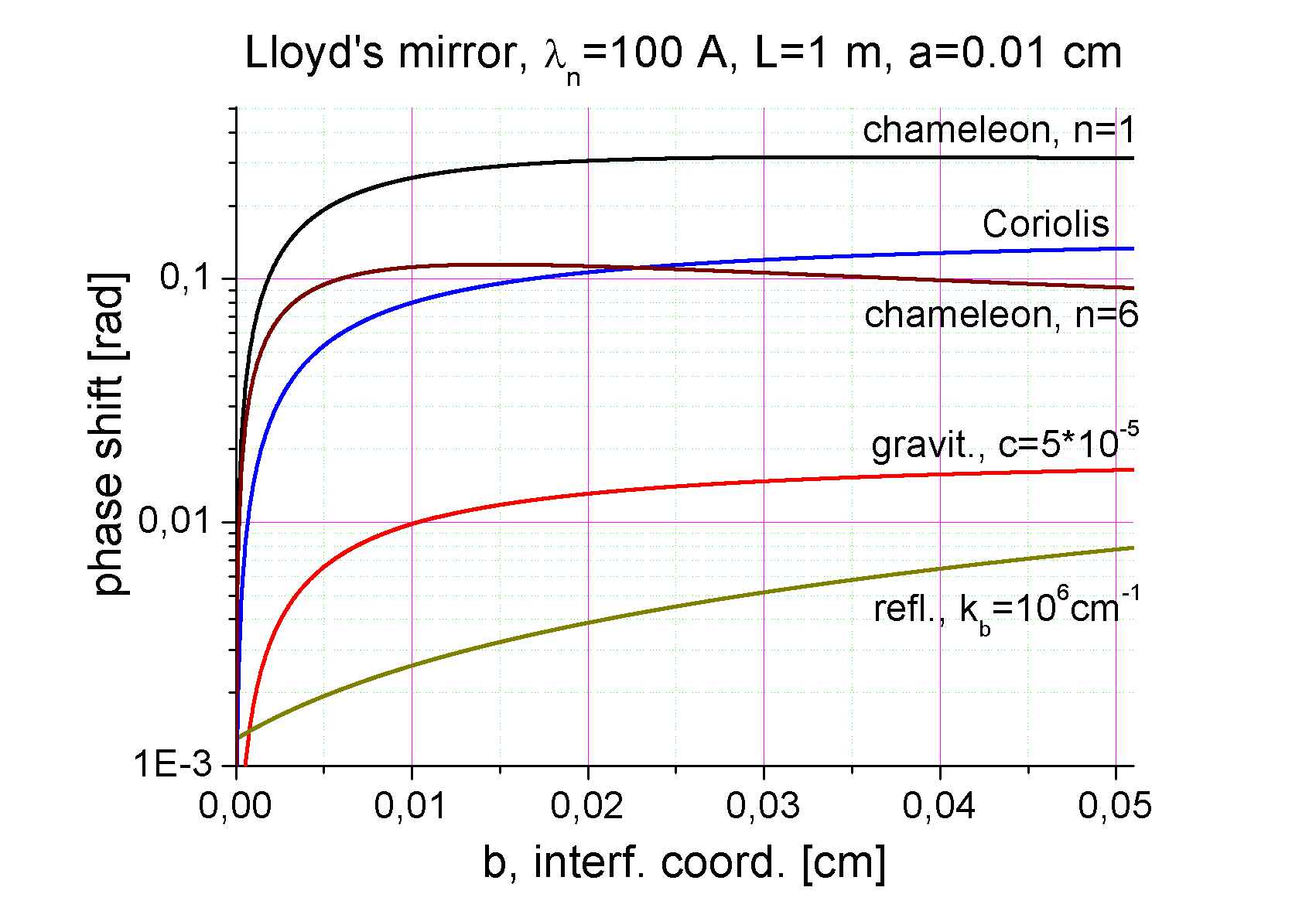}}
\end{center}
\caption{The neutron wave phase shifts $\varphi$ in the Lloyd's mirror interferometer with parameters: the
neutron wave length 100\,\AA, $L$=1 m, $a=0.01$\,cm, the interaction parameters of the chameleon field with
matter $\beta=10^{7}$,\, $n=1$ and $n=6$. Also shown: the gravitational phase shift $\varphi_{gr}$ at $c=5\times
10^{-5}$; the Coriolis phase shift, and the effect of reflection as $\delta\varphi_{refl}=\pi - \varphi_{refl}$
($k_{b}=10^{6}$\,cm$^{-1}$).}
\end{figure}

\begin{figure}
\begin{center}
\resizebox{16cm}{13cm}{\includegraphics[width=\columnwidth]{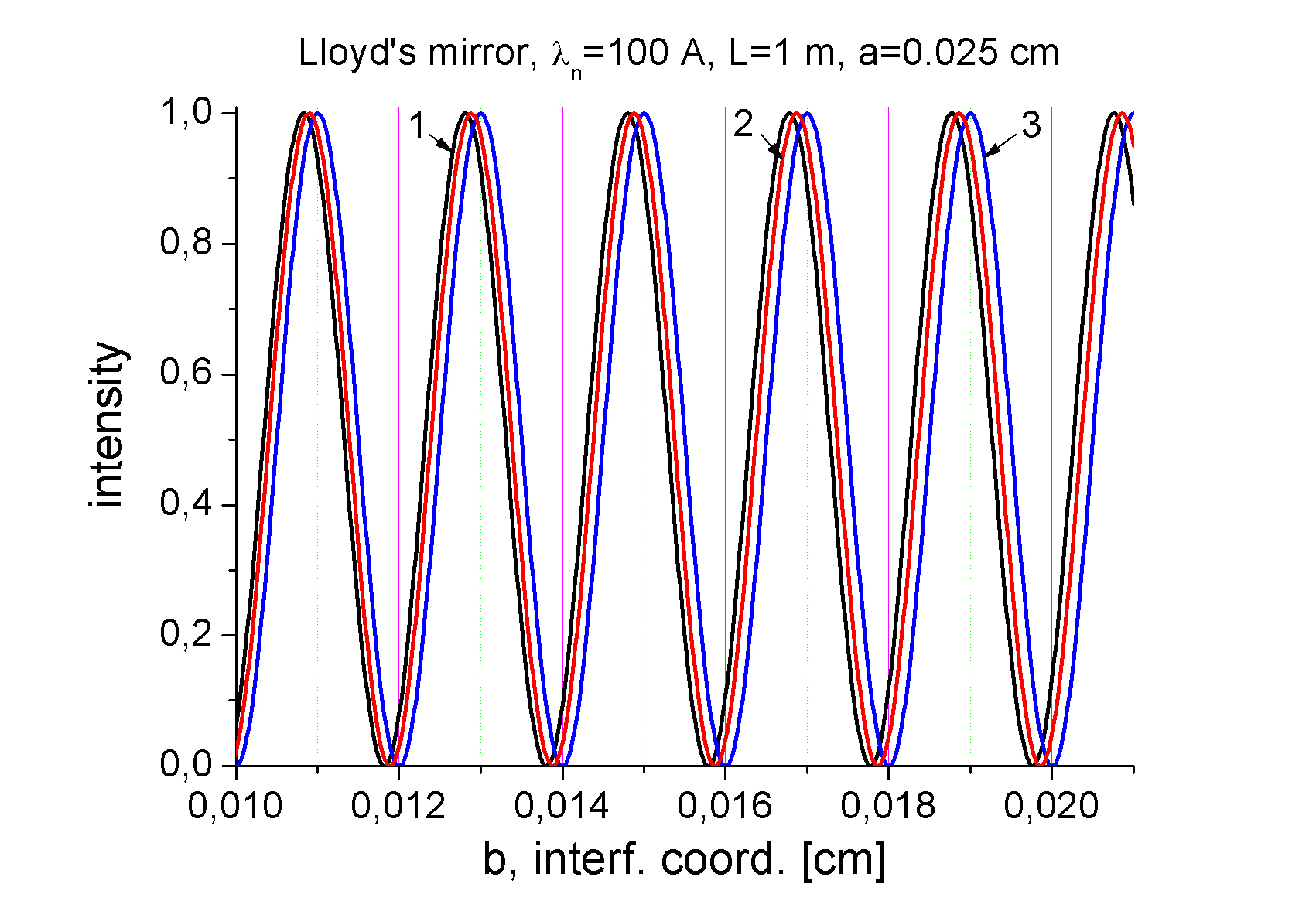}}
\end{center}
\caption{The calculated interference pattern.
 The interferometer has parameters: the neutron wave length 100\,\AA \,
$L$=1 m, $a=0.01$ \,cm. 1 - all the phase shifts are taken into account: of the chameleon field with matter
interaction parameters $\beta=10^{7}$ and $n=1$, the gravitational phase shift $\varphi_{gr}$ at $c=5\times
10^{-5}$, the Coriolis phase shift, and the phase shift of the ray II at reflection $\delta\varphi_{refl}=\pi -
\varphi_{refl}$ ($k_{b}=10^{6}$\,cm$^{-1}$); 2 - the same, but excluding all the phase shifts except the
chameleon induced phase shift; 3 - all phase shifts excluded except purely geometrical.}
\end{figure}


\begin{thebibliography}{92}
\bibitem{Ratra}
B. Ratra and P.J.E. Peebles, Phys. Rev. {\bf D37} (1988) 3406.

\bibitem{Pee}
P.J.E. Peebles and B. Ratra, Rev. Mod. Phys. {\bf 75} (2003) 559.

\bibitem{Cop}
E.J. Copeland, M. Sami, and S Tsujikawa, Int. Journ. Mod. Phys., {\bf D15}
(2006) 1553.

\bibitem{Kho}
J. Khoury and A. Weltman, Phys. Rev. Lett., {\bf 93} (2004) 171104;
Phys. Rev. {\bf D69} (2004) 044026.

\bibitem{Bra04}
P. Brax, C. van de Bruck, A.-C. Davis, J. Khoury, and A. Weltman,
Phys. Rev. {\bf D70} (2004) 123518.

\bibitem{Gub}
S.S. Gubser and J. Khoury, Phys. Rev. {\bf D70} (2004) 104001.

\bibitem{Upa06}
A. Upadhye, S.S. Gubser, and J. Khoury, Phys. Rev. {\bf D74} (2006) 104024.

\bibitem{Mota06}
D.F. Mota and D.S. Shaw, Phys. Rev. Lett. {\bf 97} (2006) 151102.

\bibitem{Mota07}
D.F. Mota and D.S. Shaw, Phys. Rev. {\bf D75} (2007) 063501.

\bibitem{Fisch}
E. Fischbach and C.L. Talmadge, {\it The Search for Non-Newtonian Gravity}
(Springer-Verlag, New-York, 1998).

\bibitem{Rep}
M. Bordag, U. Mohideen and V.M. Mostepanenko, Phys. Rep. {\bf 353}, 1
(2001).

\bibitem{Cas}
P. Brax, C. van de Bruck, A.C. Davis, D.F. Mota, and J. Shaw,
Phys. Rev. {\bf D76} (2007) 124034.

\bibitem{Cas1}
P. Brax, C. van de Bruck, A.C. Davis, D. J. Shaw, and D. Ianuzzi,
Phys. Rev. Lett. {\bf 104} (2010) 241101.

\bibitem{myN}
Yu.N. Pokotilovski, Jad. Fiz. {\bf 69} (2006) 953; Phys. At. Nucl. {\bf 69}
(2006) 924.

\bibitem{NesvN}
V.V. Nesvizhevski, G. Pignol, K.V. Protasov,  Phys. Rev. {\bf D77} (2008) 034020.

\bibitem{BraPi}
P. Brax, G. Pignol, Phys. Rev. Lett. {\bf 107} (2011) 111301.

\bibitem{Ahl}
M. Ahlers, A. Lindner, A. Ringwald, L. Schrempp, and C. Weniger,
Phys. Rev. {\bf D77} (2008) 015018.

\bibitem{Gies}
H. Gies, D.F. Mota, and D.S. Shaw, Phys. Rev. {\bf D77} (2008) 025016.

\bibitem{CHASE}
J.H. Steffen, A. Upadhue, A. Baumbaugh, A.S. Chou, P.O. Mazur, R. Tomlin,
A. Weltman,  and A. Wester, Phys. Rev. Lett., {\bf 105} (2010) 261803.

\bibitem{CHASE1}
A. Upadhue, J.H. Steffen, and A. Chou, arXiv:1204.5476 [hep-ph].

\bibitem{ADMX}
G. Rybka, M. Hotz, L.J. Rosenberg, S.J. Asztalos, G. Carosi, C. Hagmann,
D. Kinton, K. van Bibber, J. Hoskins, C. Martin, P. Sikivie, D.B. Tanner,
R. Bradley, J. Clarke, Phys. Rev. Lett., {\bf 105} (2010) 051801.

\bibitem{Cori}
S.A. Werner, J.-L. Staudenmann, and R. Colella, Phys. Rev. Lett. {\bf 42} (1979) 1103.

D.K Atwood, M.A. Horne, C.G. Shull, and J. Arthur , Phys. Rev. Lett.
{\bf 52} (1984) 1673.

%
\bibitem{Ste}
A. Steyerl, H. Nagel, F.-X. Schreiber, K.-A. Steinhauser, R. Gahler,
W. Gl\"aser, P. Ageron, J.-M. Astruc, N. Drexel, R. Gervais, W. Mampe,
Phys. Lett., {\bf A116} (1986) 347.

\bibitem{eldet}
J. Jacubek, P. Schmidt-Wellenburg, P. Geltenbort, M. Platkevic,
C. Plonka-Spehr, J. Solc, T. Soldner, Nucl. Instr. Meth., {\bf A600} (2009) 651.

T. Sanuki, S. Kamamiya, S. Kawasaki, S. Sonoda, Nucl. Instr. Meth., {\bf A600} (2009) 657.

J. Jacubek, M. Platkevic, P. Schmidt-Wellenburg, P. Geltenbort,
C. Plonka-Spehr, M Daum, Nucl. Instr. Meth., {\bf A607} (2009) 45.

S. Kawasaki, G. Ichikawa, M. Hino, Y. Kamiya, M. Kitaguchi,  S. Kawamiya,
T. Sanuki, S. Sonoda, Nucl. Instr. Meth., {\bf A615} (2010) 42.

\bibitem{emdet}
V.V. Nesvizhevsky, A.K. Petukhov, H.G. B\"orner, T.A. Baranova, A.M. Gagarski, G.A. Petrov, K.V. Protasov, A.Yu.
Voronin, S. Bae\ss ler, H. Abele, A. Westphal, L. Lucovac, Eur. Phys. Journ., {\bf C40} (2005) 479.

%
\bibitem{Dre}
W. Drexel, Neutron News, {\bf 1} (1990) 1.

%
\bibitem{BH}
U. Bonse and M. Hart, Appl. Phys. Lett. {\bf 6} (1965) 155.

%
\bibitem{RTB}
H. Rauch, W. Treimer and U. Bonse , Phys. Lett. {\bf 47A} (1974) 369.

%
\bibitem{Io}
A.I. Ioffe, Nucl. Instr. Meth., {A268} (1988) 169.

%
\bibitem{Eder}
K. Eder, M. Gruber, A. Zeilinger, P. Geltenbort, R. G\"ahler, W. Mampe, and
W. Drexel,, Nucl. Instr. Meth., {A284} (1989) 171.

%
\bibitem{Zo}
G. van der Zouw, M. Weber, J. Felber, R. G\"ahler, P. Geltenbort, A. Zeilinger,
Nucl. Instr. Meth., {A440} (2000) 568.

\bibitem{corr1}
E.H. Cook-Yahrborough, {\it Instrumentation techniques in nuclear pulse
analysis}, Washington, 1964, 207.

\bibitem{corr2}
A.I. Mogilner, O.A. Sal'nikov, L.A. Timokhin, Pribory i Tekhnika Experimenta,
No.2 (1966) 22.

\bibitem{corr3}
M.I. Novopoltsev, Yu.N. Pokotilovski, Pribory i Tekhnika Experimenta, No.5
(2010) 19; Instr. and Experim. Tekhn., {\bf 53} No.5 (2010) 635;
arXiv:1008.1419[nucl-ex].

\bibitem{res}
T. Jenke, P. Geltenbort, H. Lemmel, and H. Abele, Nature Physics, {\bf 7} (2011) 468.

\bibitem{ax-Lloyd}
Yu.N. Pokotilovski, Pis'ma v ZhETF {\bf 94} (2011) 447;
JETP Lett., {\bf 94} (2011) 413; arXiv:1107.1481 [nucl-ex].

%
\bibitem{Newt}
Yu.N. Pokotilovski, in {\it "20th Intern. Seminar on Interactions of Neutrons
with Nuclei", Alushta, Ukraine, May 21-26, 2012.},
http://isinn.jinr.ru/20/prog.html

\end{thebibliography}
\end{document}